\documentclass[twocolumn]{article}
\usepackage{graphicx}
\usepackage{amsmath}
\usepackage{array}
\usepackage{color}
\usepackage{amssymb}
\usepackage{hyperref}
\usepackage{flushend}
\usepackage{float}
\usepackage{times}
\usepackage[hang,flushmargin]{footmisc} 
\usepackage[margin=2cm]{geometry}
\usepackage{natbib}
\usepackage{afterpage}
\begin{document}
%\firstpage{1}

%\title[Modeling DNA Methylation Dynamics]{Phylo\hspace{1pt}(epi)\hspace{1pt}genetic modeling of DNA methylation data}
%\history{\phantom{x}  }
%\editor{\phantom{x}  }

\twocolumn[
  \begin{@twocolumnfalse}

	\title{Modeling DNA methylation dynamics with approaches from phylogenetics}
\author{John A. Capra\,$^{1,*}$ and Dennis Kostka\,$^{2,}$\footnote{to whom correspondence should be addressed}}

	\maketitle

\begin{abstract}
\noindent \emph{\bf Motivation:} 
Methylation of CpG dinucleotides is a prevalent epigenetic modification that is required for proper development in vertebrates, and changes in CpG methylation are essential to cellular differentiation. Genome-wide DNA methylation assays have become increasingly common, and recently distinct stages across differentiating cellular lineages have been assayed. However, current methods for modeling methylation dynamics do not account for the dependency structure between precursor and dependent cell types.\vspace{.5ex}

\noindent\emph{\bf Results:}  We developed a continuous-time Markov chain approach, based on the observation that changes in methylation state over tissue differentiation can be modeled similarly to DNA nucleotide changes over evolutionary time. 
This model explicitly takes precursor to descendant relationships into account and enables inference of CpG methylation dynamics. To illustrate our method, we analyzed a
high-resolution methylation map of the differentiation of mouse stem cells into several blood cell types. Our model can successfully infer unobserved CpG methylation states from observations at the same sites in related cell types (90\% correct), and this approach more accurately reconstructs missing data than imputation based on neighboring CpGs (84\% correct). Additionally, the single CpG resolution of our methylation dynamics estimates enabled us to show that DNA sequence context of CpG sites is informative about methylation dynamics across tissue differentiation. Finally, we identified genomic regions with clusters of highly dynamic CpGs and present a likely functional example. 
Our work establishes a framework for inference and modeling that is well-suited to DNA methylation data, and our success suggests that other methods for analyzing DNA nucleotide substitutions will also translate to the modeling of epigenetic phenomena. \vspace{.5ex}

\noindent \emph{\bf Availability:} Source code and inferred methylation variation rate categories are available upon request.\vspace{.5ex}

\noindent \emph{\bf Contact:} \href{tony.capra@vanderbilt.edu}{tony.capra@vanderbilt.edu}, \href{kostka@pitt.edu}{kostka@pitt.edu}
\vspace{1cm}

\end{abstract}

\end{@twocolumnfalse}
]

\let\thefootnote\relax\footnotetext{ $^1$ Center for Human Genetics Research and Department of Biomedical Informatics, Vanderbilt University, Nashville, TN, 37232\\
$^{2}$Departments of Developmental Biology and Computational \& Systems Biology, University of Pittsburgh, Pittsburgh, PA, 15201\\
$^*$ Corresponding author}

\section{Introduction}

DNA methylation is a common epigenetic modification essential to organism development~\citep{smith13}. In vertebrates, DNA is most commonly methylated at the fifth carbon position on cytosine nucleotides (5mC) that are followed by a guanine, so-called CpG sites. A family of three DNA methyltransferase enzymes (DNMT1, DNMT3A, DNMT3B) is responsible for the establishment and maintenance of methylation state at the millions of CpG sites in most mammalian genomes~\citep{smith13}.
Recently the ability to perform genome-wide assays of the methylation state of individual CpGs has become a reality due to advances in microarray and DNA sequencing technology. Several approaches that vary in their accuracy, biases, coverage, and cost are commonly used; see~\citet{laird10} for a detailed review of current methods.

Systematic screening of DNA methylation across tissue diff\-er\-ent\-iation and develop\-ment has improved our knowledge of its role in these processes~\citep{bock12,xie13}. The methylation profile of the mammalian genome is largely stable, but the methylation of specific genomic regions changes dynamically across development, and different cellular lineages have unique methylation profiles~\citep{ziller13}. Additionally, the DNA methylation state nearby a gene's transcription start site correlates with gene expression \citep{xie13}, and the correct orchestration of methylation changes is essential for proper cellular differentiation. Aberrant methylation changes may lead to tumorgenesis and other diseases~\citep{portela10, tost10,hansen2011increased,bergman13}. 

\begin{figure*}[!tpb]
%%%%%%%%%%%%%%%%%%%%%%%%%%%%
\begin{minipage}{.41\textwidth}
\caption{CpG methylation dynamics can be modeled with an approach inspired by phylogenetic analysis of nucleotide substitutions. 
Left: A lineage tree of sampled blood cell types during hematopoietic differentiation with stem cells on top and terminally differentiated cells on the bottom (see Section \ref{sec:datexplain} for details).  The lineage tree takes the role played by the species tree in the phylogenetic context.
Right:  Examples of methylation patterns across differentiation (columns) for CpG sites at different genomic locations. Each row corresponds to a cell type in the linage tree on the left. The white block represents missing data. The discretized methylation states are analogous to DNA sequence data.}\label{fig:explanation}
\end{minipage}
\hfill
\begin{minipage}{.57\textwidth}
\flushright
\includegraphics[width=.99\textwidth]{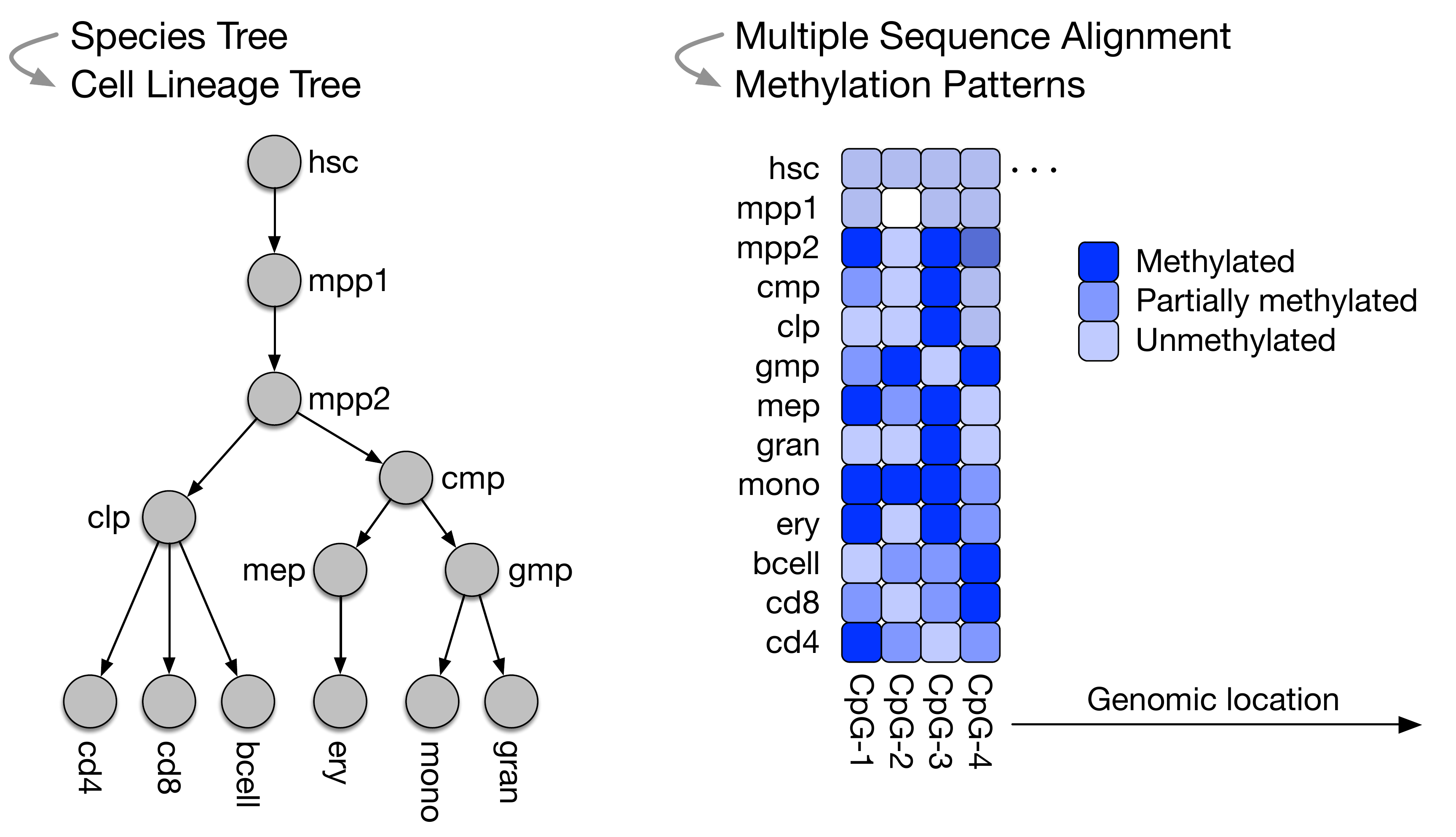}
\end{minipage}
\end{figure*}

Studies assaying DNA methylation often focus on the comparison of two types of conditions, like tumor vs.~normal tissue~\citep{nordlund13}, or stem cells vs.~lineage-committed cells \citep{xie13}. However, the natural process of cellular differentiation and development has an essentially tree-like topology, in which precursor cell types are connected to their descendants by edges, thereby forming a so-called \emph{lineage tree}~\citep{frumkin05}. For example, Figure~\ref{fig:explanation} depicts a lineage tree for blood cell differentiation, where DNA methylation has been assayed in cell types represented by nodes~\citep{bock12}. 
Independent pairwise comparisons cannot accommodate this structure. 

To address this issue, we introduce an approach to model methylation state changes between cell types that explicitly takes dependencies induced by the lineage tree into account.
In this setup, modeling methylation changes over developmental time is in many ways reminiscent of describing DNA nucleotide changes over evolutionary time (Figure \ref{fig:explanation}). As a result, we adapt established continuous-time Markov models of sequence evolution to fit this task. 
In addition to accommodating cell lineage relations during de\-vel\-op\-ment, our approach has the  benefit that it works at single CpG dinucleotide resolution and does not require the spatial aggregation of methylation measurements across the genome. 
Finally, the analogy with models for DNA sequence evolution provides intuitive means to handle missing data, which are common in many DNA methylation datasets. During parameter estimation, missing data can be marginalized over, and the equivalent of joint ancestry reconstruction~\citep{pupko00} allows efficient inference of the most likely methylation states for unobserved data in context of the lineage tree.

As an illustration of our approach, we analyzed methylation data collected across the cell lineage tree in Figure \ref{fig:explanation} from~\citet{bock12}. On this dataset, our method enabled the accurate reconstruction of missing methylation states.  The single CpG resolution of our analysis allowed us to discover that the identity of neighboring dinucleotides is strongly correlated with CpG methylation dynamics at many sites the mouse genome. Finally, using our predictions of CpG methylation variability, we identified a cluster of highly dynamic CpG sites that show evidence of enhancer activity in blood cells.
%

%========================================================================
%\begin{methods}
\section{Methods}
%========================================================================

%================================================================
% FIGURE 1
%================================================================

\subsection{Modeling methylation changes across tissue differentiation}
%======================================================================

We model the dynamics of DNA methylation across cellular differentiation using an approach motivated by phylogenetic models. In the phylogenetic context, a continuous time Markov chain is used to quantify DNA sequence changes between species over a known species tree. Intuitively, we adapt this approach and replace the species tree with a cell lineage tree and the four-state alphabet of DNA with a three-state alphabet based on methylation status.  In our model, the cell lineage tree consists of nodes that correspond to cell types and edges

In order to model transitions between CpG methylation states along edges of the cell lineage tree, we associate each node  with a discrete random variable $X_{i}$ ($1\leq i \leq N$, assuming $N$ nodes), which is dependent on its parents (i.e., its direct precursor cell types) in the lineage tree. As in DNA sequence evolution models, we use a continuous time Markov chain to describe this dependency structure. Specifically, if nodes $i$ and $j$ are connected in the lineage tree by an edge $i \rightarrow j$ of length $t$, then the probability of the methylation state at node $j$ being $l$ conditional on node $i$ being in methylation state $k$ is given by
\[
\label{eq:transition}
\phantom{\prod_1^2}P(X_j=l|X_i=k) = [\mbox{expm}(Qt)]_{kl}
\]
\noindent for $k,l \in \{{\tt u},{\tt p},{\tt m}\}$~\citep{guttorp95} . $Q$ is a $3\times 3$ rate matrix (or generator), and $\mbox{expm}$ denotes the matrix exponential. We assume a time-reversible Markov chain with equilibrium frequency $\pi$, which implies $Q$ is fully parameterized by three non-negative rate-parameters, $\{a_i\}$, and $\pi$.\footnote{The number of expected transitions along an edge is $(-1)\sum_i\pi_i Q_{ii}t$, and therefore we will enforce $(-1)\sum_i\pi_iQ_{ii} =1$ and report $t$ in units of expected methylation state transitions.}
In summary, our model is parameterized by $\vartheta$, which consists of: the topology of the lineage tree (which we assume is fixed, known, and consists of $N$ nodes and $E$ edges), the branch lengths $\{t_i\}_{i=1}^E$, the equilibrium frequencies $\{\pi_i\}_{i=1}^3$, and the rate parameters $\{a_i\}_{i=1}^3$. The likelihood of an observed methylation pattern $\mathbf{x} = \{x_i\}_{i=1}^N$ is then
\[
P(\mathbf{x}|\vartheta) = \prod_{i=1}^N P_{\vartheta}(X_i=x_i|X_{{\tt pa}(i)} = x_{{\tt pa}(i)})
\]
\noindent where ${\tt pa}(i)$ is the parent of node $i$ in the lineage tree.
For the root node we have $P(X=i) = \pi_i$, and assuming independence between methylation patterns at different CpG sites, we have for the likelihood of all observed patterns $D =\{\mathbf{x}_i\}_{i=1}^L$ (assuming there are $L$ CpG sites):
\begin{equation}
\label{eq:likelihood}
L(\vartheta ) = P(D|\vartheta ) = \prod_i^L P(\mathbf{x}_i|\vartheta).
\end{equation}
In contrast to most applications dealing with DNA sequence changes, non-leaf nodes can be observed in our setting. We handle missing data by marginalization, i.e., summation over all possible configurations of un-observed nodes in the lineage tree, which can be done efficiently via the elimination algorithm~\citep{siepel05}.
Maximum likelihood parameter estimates are then obtained by maximizing Equation (\ref{eq:likelihood}) over branch lengths, equilibrium frequencies and rate parameters.
In summary, we have adapted a well-know class of models that is typically used in the context of DNA sequence evolution to model the dynamics of methylation changes during tissue differentiation.

\subsection{Integrating rate heterogeneity}
%==========================================

\subsubsection{Modeling rate heterogeniety} \label{sec:rateHetero} The approach described so far models methylation dynamics using the same process at all CpG sites in the genome, and thereby assumes homogeneity of methylation dynamics. This assumption is not always reasonable. For instance, CpGs located in CpG islands have a propensity to be unmethylated (compared to other CpG sites), and a disposition to stay in that state \citep{jones12}.

To address this issue, we incorporate rate heterogeneity into our model using a mixture modeling approach similar to phylogenetic models for DNA changes under heterogeneous substitution rates. First, we assume a certain fraction ($\beta$) of CpG sites to be \emph{invariant}, i.e.,~they do not change their methylation state during tissue differentiation. For the $(1-\beta)$ fraction of \emph{variable} CpG sites, we assume $M$ different equiprobable rate categories $\{r_m\}_{m=1}^M$ such that the probability of methylation pattern $\mathbf{x}_i$ is now
\[
P(\mathbf{x}_i|\tilde{\vartheta}) = \beta P(\mathbf{x}_i|r_0,\vartheta) + (1-\beta) \frac{1}{M} \sum_{m=1}^M P(\mathbf{x}_i|r_m,\vartheta),
\]
where we have used the ``rate category'' $r_0$ to denote invariance and $\tilde{\vartheta}$ to denote the new parameter set.
For the invariant term on right hand side above, $P(\mathbf{x}_i|r_0,\vartheta) = p_k$ if all methylation states in $\mathbf{x}_i$ are $k$ (for $k \in \{{\tt u},{\tt p},{\tt m}\}$) and zero otherwise. For the variable part, we have $P(\mathbf{x}_i|r_m,\vartheta) = P(\mathbf{x}_i|\bar{\vartheta}(m))$, where we use Equation (\ref{eq:likelihood}), but with all branch lengths in $\vartheta$ scaled by the factor $r_m$. The scale factors $\{r_m\}$ are determined by a Gamma distribution with shape parameter $\alpha$ and scale parameter $1/\alpha$ (setting the scale parameter to $1/\alpha$ ensures that the Gamma distribution has a mean of one). Next, the probability density function of the Gamma distribution is discretized by splitting its domain into $M$ equal-mass bins and setting $r_m$ equal to the mean conditional on bin $m$. Thus, a single positive parameter $\alpha$ determines all $M$ rates. The additional parameters to account for rate variation between CpG sites are: the fraction of invariant sites $\beta$, the frequencies of invariant states $\{p_i\}_{i=1}^3$, and the shape parameter $\alpha$ of the Gamma distribution. Maximum likelihood estimates are again obtained considering CpG sites as independent. In summary, we use the $\Gamma + I$ model to account for rate heterogeneity across different CpG sites.

\subsubsection{Assigning CpG sites to rate categories:} To assign CpG sites to rate categories we use an empirical Bayes approach \citep{galtier05}. Let $\hat{\vartheta}$ denote the maximum likelihood estimates for $\tilde{\vartheta}$. We assign methylation pattern $\mathbf{x}_i$ to rate category $\hat{m} = \mbox{argmax}_m P(\mathbf{x}_i | \hat{r}_m , \hat{\vartheta}) P(m) / P(\mathbf{x}_i |\hat{\vartheta})$, where $P(m) = \hat{\beta}$ for $m=0$ and $P(m) = (1-\hat{\beta})/M$ for $1\leq m \leq M$.

\begin{figure*}[htb]
\begin{minipage}{.59\textwidth}
\centerline{\includegraphics[width=1\textwidth]{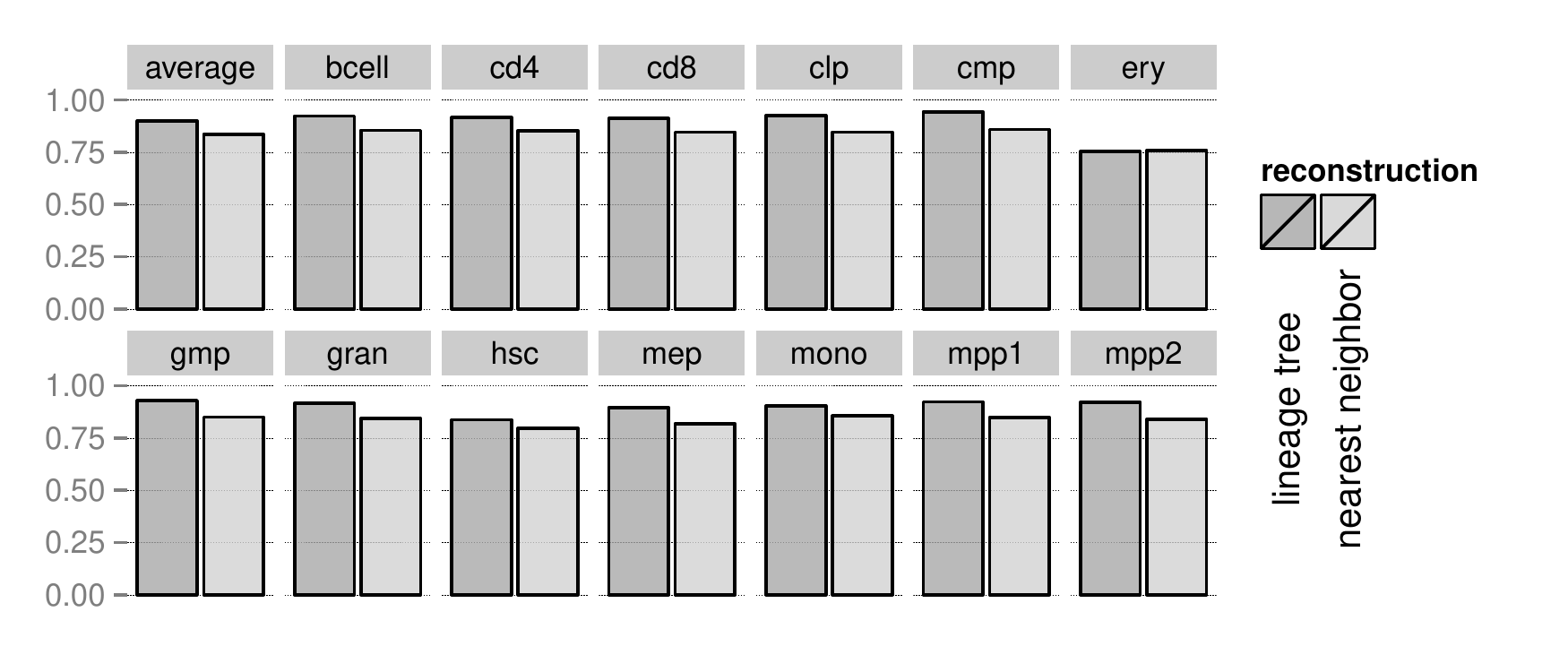}}
\end{minipage}
\hfill
\begin{minipage}{.4\textwidth}
\caption{Lineage-tree-based reconstruction of methylation state is more accurate than nearest-neighbor-based reconstruction. We masked the methylation status for 10,000 CpG sites in each cell type and reconstructed these values using ``vertical'' information from our lineage tree model and ``horizontal'' information from neighboring CpG sites. The lineage tree approach proved significantly more accurate overall (90\% vs. 84\%) and within every cell type except erythrocytes, which have the longest branch length.}\label{fig:errors}
\end{minipage}
\end{figure*}

\subsection{Reconstructing missing data}
%=======================================

\label{sec:ancML}

Our model, which we refer to as a ``phylo-epigenetic'' model, can reconstruct missing or unobserved methylation states a cell type. Intuitively, for a given CpG site, nearby cell types in the lineage tree carry information about its likely methylation state. We quantify this relationship using joint maximum likelihood  ancestry reconstruction \citep{pupko00}. In essence, assume methylation pattern $\mathbf{x}_i$ contains one or more missing values (i.e., unobserved methylation states). Further assume the empirical Bayes procedure discussed above assigns pattern $\mathbf{x}_i$ to rate category $m$. Note that during this procedure missing values in $\mathbf{x}_i$ had been ``marginalized out''.
Then, we assign the missing values in $\mathbf{x}_i$ to the methylation state configuration that maximizes the likelihood $P(\mathbf{x}_i|\hat{r}_m,\hat{\vartheta})$. This can be efficiently computed, also for multiple missing values at the same site \citep{pupko00}.

This reconstruction strategy shares methylation state information for a CpG site ``vertically'' across the lineage tree, and is complementary to approaches leveraging ``horizontal'' correlations between different but nearby CpG sites across the genome.

\subsection{Data sources and processing} \label{sec:datexplain}
%=================================================================================================

We analyzed DNA methylation maps from 13 cell populations from stages of a differentiation of adult mouse hematopoietic stem cells to different blood lineages, which are available in the Supplementary Material of~\citet{bock12}. The purified cell types were obtained at progressive levels of differentiation, starting with hematopoietic stem cells (HSC), followed by multipotent progenitor cells (MPP1 and MPP2), and progenitor cells of the lymphoid (CLP) and myeloid (CMP) lineages. For the lymphoid progenitors, further differentiated cells included T helper cells (CD4), T cells (CD8) and B cells (BCELL). For myeloid progenitor cells, the next stages were  granulocyte-monocyte progenitors (GMP) and megakaryocyte-erythroid progenitors (MEP); the former was followed by monocytes (MONO) and granulocytes (GRAN), whereas the latter was followed by  erythrocytes (ERY). The relationships between cell types are summarized in the lineage tree in Figure \ref{fig:explanation}. \citet{bock12} generated a methylation map for each cell type using reduced representation bisulfite sequencing (RRBS). 
We averaged the two RRBS replicates for each cell type, and for each CpG site with RRBS data, we discretized the methylation status into methylated ($>$ 0.8 methylated), partially methylated (between 0.1 and 0.8), and unmethylated ($<$ 0.1) categories based on the fraction of methylated reads for the site. Histograms of these values showed clear peaks at the ends of the spectrum.
We defined CpG islands using the {\tt cpgIslandExt} table for the {\tt mm9} build of the mouse gene from the UCSC Genome Browser~\citep{kent02}. 
To estimate the parameters of the model described in Section \ref{sec:rateHetero} we used the observed frequencies (excluding CpGs with missing values) for $\{p_i\}_{i=1}^3$, and used the L-BFGS-B algorithm to obtain maximum likelihood estimates for all other parameters. 

%\end{methods}

%===========================================================
\section{Results}
%===========================================================

We applied our phylo-epigenetic model to an RRBS data set tracing the differentiation of adult hematopietic stem cells \citep{bock12}. This study queried a set of over two million CpG sites at different stages during blood lineage differentiation, and the  assayed cell types with their relationships are summarized in the lineage tree in Figure \ref{fig:explanation}. 

We fit a phylo-epigenetic model with four rate categories (three variable and one invariant) to the discretized methylation status of CpG sites along each chromosome (Methods). The maximum likelihood estimates of the model parameters qualitatively agree across chromosomes, and the resulting model is consistent with several previous findings. Invariant CpG sites are more prevalent than variable sites~\citep{bock12,ziller13}; 62\% of CpGs are invariant in our analysis, and 15\%, 9\%, and 14\% fall into the slow, medium, and fast rate categories, respectively. As expected, the equilibrium distribution for variable states favors methylated CpGs (i.e., $\hat{\pi}_{\tt u} < \hat{\pi}_{\tt p} < \hat{\pi}_{\tt m}$ for all chromosomes). Contrasting the dynamics at variable and invariant CpG sites, we find that invariant sites are most likely to remain unmethylated during differentiation (61\%), whereas variable sites are most likely to be in methylated states (57\%). 
The branch lengths obtained in our fitted models reflect the number of expected methylation state transitions between cell types, and we see the longest branch lengths between MEP and ERY cells and between HSC and MPP1 cells ($t=5.89$ and $t=0.98$ averaged over chromosomes, respectively).  
These numbers correspond to an expected fraction of CpG sites with \emph{observed} methylation changes of 23\% for MEP$\rightarrow$ERY and  of 15\% and for HSC$\rightarrow$MPP1, taking into account the prevalence of invariant sites and the different rate categories for variable sites.

In the next three sections, we present examples of how our modeling approach enables analysis of methylation dynamics across cellular differentiations.

\begin{figure*}[htb]
\begin{minipage}{.49\textwidth}
\centerline{\includegraphics[width=.99\textwidth]{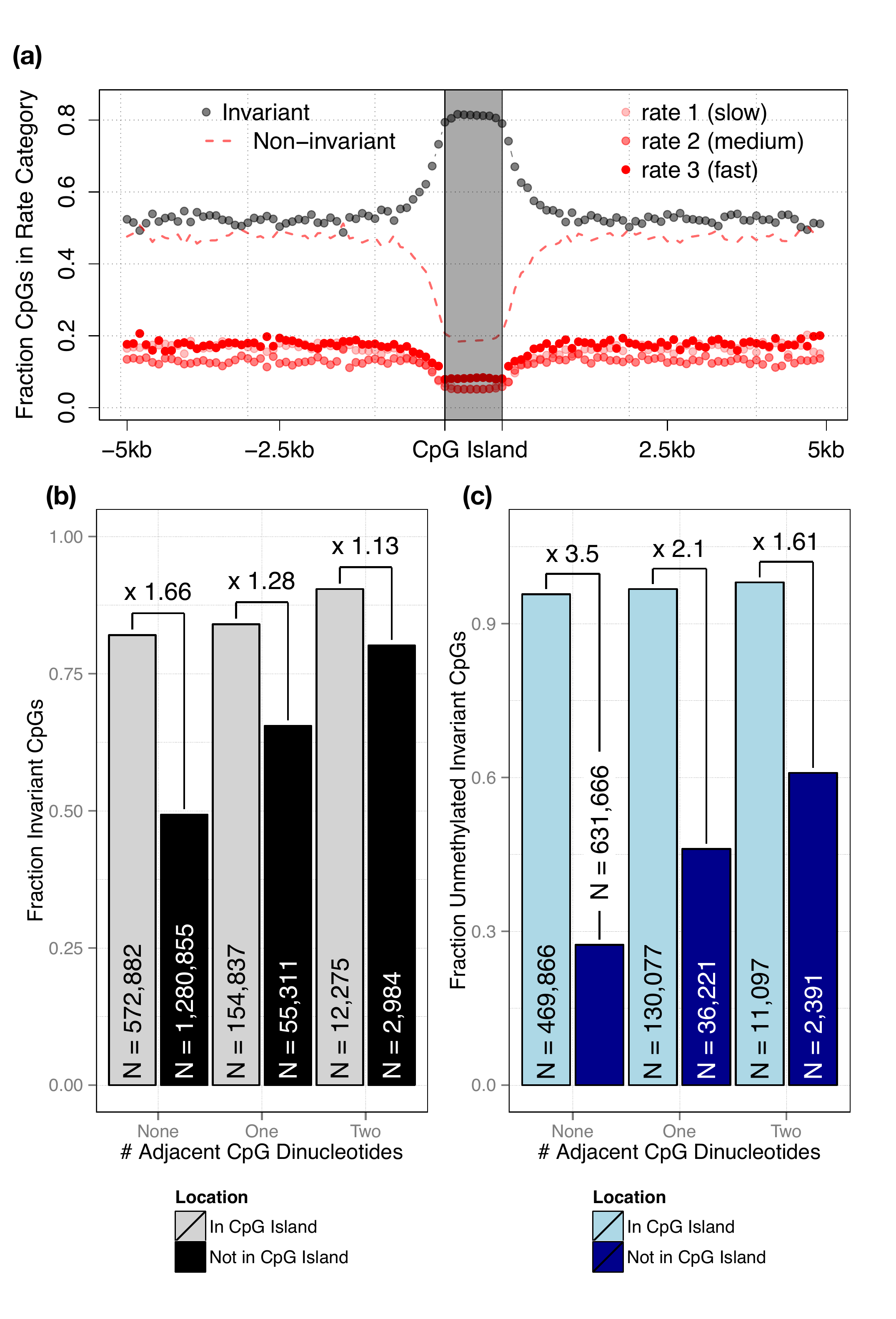}}
\end{minipage}
\hfill
\begin{minipage}{.49\textwidth}
\caption{Methylation dynamics are influenced by DNA sequence context outside of CpG islands. \textbf{(a)} Invariant CpG sites (gray circles) are the most common category in our analysis, and they are strongly enriched in CpG islands. The other rate categories (red circles) are roughly equally likely in and around CpG islands. Each dot represents an average over evenly sized bins centered on the 16,024 CpG Islands. The red dashed line gives the sum of the variable categories. \textbf{(b)} The presence of adjacent CpG sites is strongly correlated with CpG methylation dynamics outside of CpG islands. The effect of DNA sequence context is much weaker within CpG islands. \textbf{(c)} Nearly all invariant CpG sites within CpG Islands are unmethylated. Outside of CpG islands, the adjacent CpG count for a site is strongly correlated with its methylation state.}\label{fig:cpg_islands}
\end{minipage}
\end{figure*}

\subsection{Reconstructing unobserved methylation states} \label{sec:reconstruction}
%========================================================

Over the 2,079,144 CpG sites assayed over 13 cellular contexts in the blood differentiation data set, 5,940,467 out of 27,028,872 (22\%) methylation states are missing, due to a range of technical issues~\citep{bock12}.  Given the prevalence of missing data, we assessed the ability of our model to reconstruct unobserved values using information from the observed methylation status of the same sites at other nodes in the lineage tree. Conceptually, this is akin to the problem of ancestral sequence reconstruction for DNA substitution models~\citep{pupko00}, but in the context of methylation, we also have observed data on internal nodes of the tree. 

For each cell type, we masked 10,000 CpG sites with measured methylation state, re-estimated model parameters on data missing the masked methylation states, reconstructed the masked values as described in the Methods, and then compared the reconstructed methylation values with the actual values. 

The reconstructed methylation states are generally very accurate (90\% correct overall), with some differences in performance between different cell types (Figure \ref{fig:errors}). As expected, the length of the edges connecting cell types is correlated with the accuracy the reconstruction of missing values; the nodes with the longest incident edges in the lineage tree (HSC and ERY) are the most difficult to reconstruct.

To compare our model's reconstruction with a baseline method, we also reconstructed the methylation state of each masked CpG based on the methylation state of its nearest neighbor (in terms of genomic location) in the same cell type. This type of reconstruction assumes a ``horizontal'' (i.e., location-wise) correlation between methylation states of neighboring CpG sites, whereas our approach can be viewed as assuming a ``vertical'' (i.e., progenitor to descendant) correlation between the same CpG site in neighboring cell types. 

Overall, lineage tree based reconstruction performs better than location-based reconstruction (Figure \ref{fig:errors}; 90\% correct vs.~84\% correct). However, we note that more sophisticated methods have achieved higher performance~\citep{zhang13} by integrating multiple complementary data types. Stratifying reconstructed methylation states by our inferred rate categories revealed that, unsurprisingly, lineage-tree-based reconstruction performs worst for ``fast'' CpG sites, i.e., those in the fastest rate category according to the empirical Bayes procedure. 

\subsection{DNA sequence context is correlated with CpG methylation dynamics.}

Having established that our approach can successfully reconstruct unobserved methylation states, we now describe several analyses that use our model's estimates of CpG methylation dynamics at single CpG resolution. 
Specifically, we show that outside CpG islands, and particularly for CpGs in promoters, the immediate DNA sequence content around CpG dinucleotides is correlated with the variability of their methylation state. 

\begin{figure*}[tpb]
\begin{minipage}{.49\textwidth}
\centerline{\includegraphics[width=2.5in]{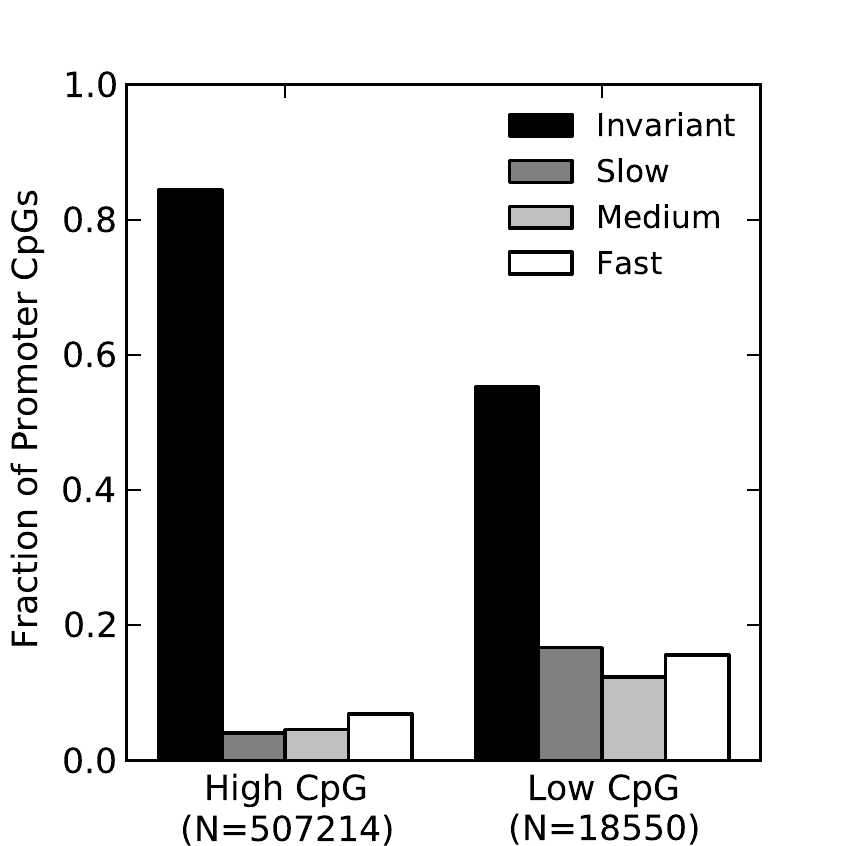}}
\end{minipage}
\hfill
\begin{minipage}{.49\textwidth}
\caption{Low CpG content promoters are enriched for CpGs with variable methylation state. We stratified mouse gene promoters into low and high CpG content groups and then compared the inferred dynamics of CpG sites in these groups. Low  CpG content promoters were significantly less likely to be in the invariant rate category ($p\approx0$; chi-squared test). This pattern remained when CpG islands were not considered. }\label{fig:promoters}
\end{minipage}
\end{figure*}

\subsubsection{Local CpG sequence context correlates with methylation dynamics outside CpG islands.}
%--------------------------------------------------------------------------------------------------
CpG islands are genomic regions with high CpG dinucleotide frequency, in which methylation status has been reported to influence gene expression \citep{illington09,jones12}. Here we study the methylation dynamics at these loci by analyzing the rate category assigned to each CpG island CpG dinucleotide for which methylation states were assayed.  

As expected, we find that invariant CpG dinucleotides are strongly enriched in CpG islands, and that this enrichment falls off with increasing distance from the island (Figure \ref{fig:cpg_islands}a). To obtain this aggregate view, we split the roughly 16,000 annnotated CpG islands into an equal number of bins and discretized flanking genomic regions into equal-sized non-overlapping tiles. The averages for corresponding locations across CpG island loci are shown as points. The strong invariance of CpG islands is in agreement with the notion that CpG islands tend to retain their methylated state \citep{jones12}. 

Taking advantage of the single CpG resolution of our approach, we explored whether the enrichment of invariant CpG dinucleotides is exclusive to CpG islands. We hypothesized that local CpG content could be important, so we stratified each CpG dicnucleotide by ($i$) whether its two neighboring dinucleotides contain none, one, or two CpGs and ($ii$) whether it is located inside a CpG island. For CpGs without neighboring CpG sites, those in CpG islands are strongly enriched for invariance over those outside of CpG islands (factor $1.66$), but this enrichment decreases for CpGs with one or two neighboring CpGs (Figure~\ref{fig:cpg_islands}b). In other words, outside of CpG islands, local CpG sequence context is strongly correlated with the absence of methylation state changes across hematopoietic differentiation.

Next, given the importance of invariant CpGs, we assessed if there is a preference for  methylated (or partially methylated) states compared to unmethylated states among invariant CpG sites. We find that invariant CpG sites in CpG islands are almost always unmethylated (90\%), which is expected from previous analyses~\citep{jones12}. However, invariant CpG sites outside of CpG islands are significantly more likely to be methylated (65\% methylated and 28\% unmethylated; $p\approx 0$, binomial test). We again hypothesized that adjacent sequence context could influence methylation at these sites. Figure \ref{fig:cpg_islands}c shows that invariant CpGs with neighboring CpG dinucleotides outside CpG islands are more unmethylated compared to their counterparts without neighboring CpG dinucleotides. 

\subsubsection{Low CpG content promoters are enriched for variable CpG sites.}

The previous subsection shows that local CpG context is associated with the dynamics of individual CpG sites in certain settings. Promoter CpGs are known to be functionally important and influenced by CpG content, so next we analyzed single CpG dynamics in promoters with respect to their overall CpG density.

The methylation state of CpGs in gene promoters is associated with transcription levels~\citep{jones12}. In several cell types, promoter methylation is negatively correlated with gene expression, and the effect is strongest in promoters with low CpG density~\citep{xie13}. The rate category assignments from our model enabled us to test whether promoter CpG content is also correlated with methylation dynamics across hematopoietic differentiation. Following~\citet{xie13}, we defined ``promoters'' as regions 500 bp upstream and downstream of transcription start sites (TSSs), and we analyzed CpG sites within this window for 19,244 mouse genes. We stratified promoters into low and high CpG density groups. As seen in human data, the CpG density distribution surrounding the mouse TSSs has two peaks; one at low ($<$0.034 CpG/bp) and one at high CpG density ($\ge$0.034 CpG/bp).

The low CpG content promoters have a significantly higher fraction of variable CpG sites compared to the high CpG promoters (Figure~\ref{fig:promoters}; $p\approx0$, chi-squared test). Nearly all (88\%) of the high CpG promoter CpG sites were invariant across the differentiation, while only 55\% of the low CpG sites were invariant. This pattern could be driven by the invariance of CpG Islands (Figure~\ref{fig:cpg_islands}) and their prevalence in high CpG content promoters, but the effect remained when CpG island sites were removed (76\% vs. 54\%; $p\approx0$). These results are consistent with the previous observation that the correlation between methylation state and gene expression is strongest in low CpG promoters~\citep{xie13}.

\begin{figure*}[htpb]
\centering
\begin{minipage}{.67\textwidth}
\centerline{\includegraphics[width=.92\textwidth]{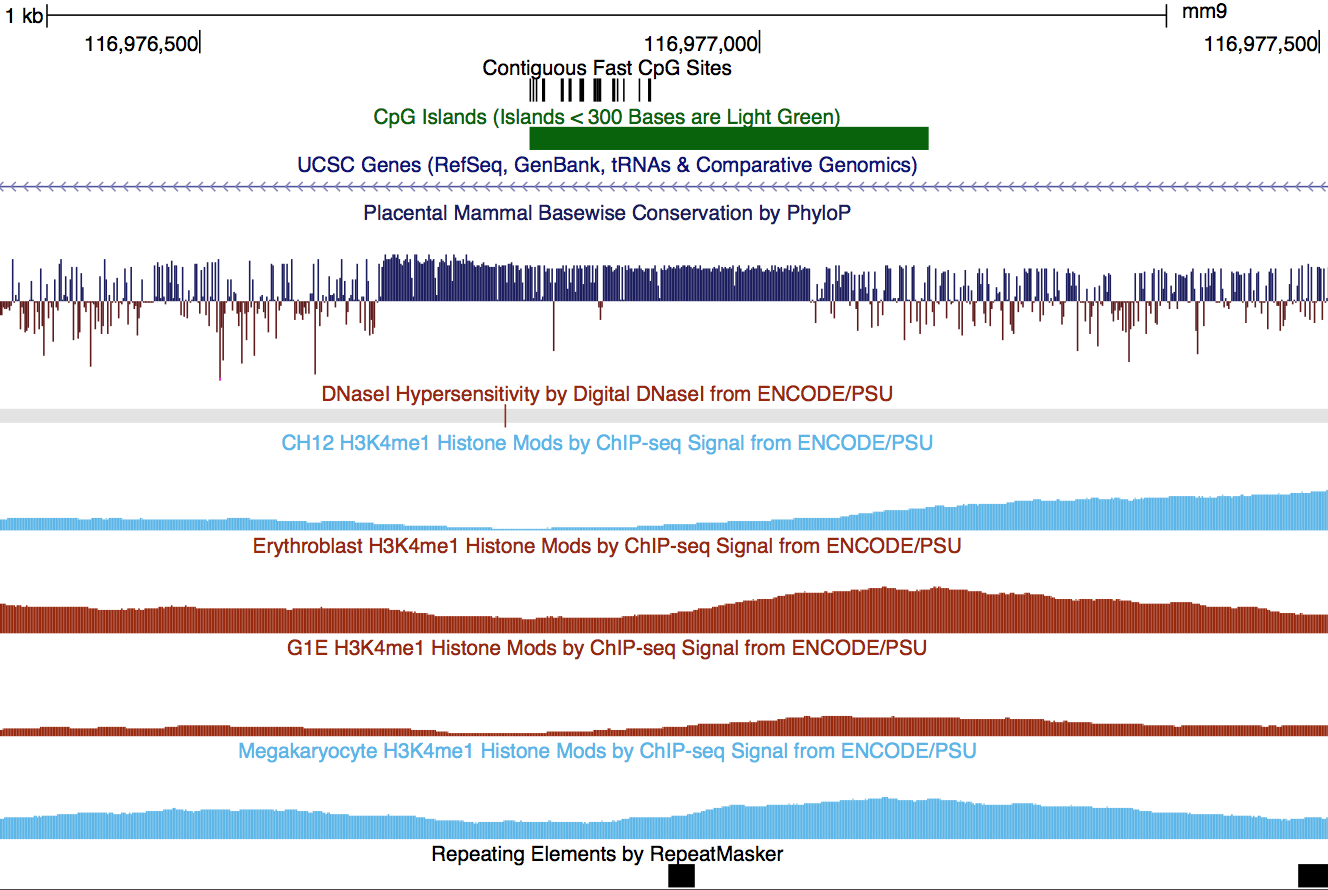}}
\end{minipage}%
\begin{minipage}{.33\textwidth}
\caption{A block of highly variable CpG sites has evidence of gene regulatory enhancer activity in several blood cells. This region on chromosome 4 (mm9.chr4:116,976,784--116,976,912) contains 16 CpG sites that our model places in the fast rate category within 109 bp. The region is located within a CpG island in an intron of the gene \textit{Rnf220}, a ubiquitin ligase. The DNA sequence at this locus is strongly conserved across placental mammals; this suggests that it is likely functionally important. In addition, functional genomics data collected by the ENCODE project~\citep{mouseENCODE} suggest that this locus is a regulatory enhancer in several blood cell types. It overlaps a DNaseI hypersensitive site in an erythroid progenitor (G1E), and it has peaks of the H3K4me1 enhancer-associated histone modification in B-cell lymphoma cells (CH12), erythroblasts, G1E cells, and megakaryocytes. \label{fig:browser} }
\end{minipage}
\end{figure*}

\subsection{Identification of genomic regions with variable methylation state}

The CpG site rate category predictions from our model enable us to identify genomic regions with frequent methylation state changes across hematopoietic differentiation. To do this, we discarded CpG dinuclotides in repeat masked regions ({\tt rmsk} track for {\tt mm9} from UCSC genome browser) and then identified maximal subsets in the RRBS data, for which each site is no more than 20 bp from the nearest other CpG in the subset. We furhter filtered out short regions ($<$50bp), and focused on regions without evidence for accelerated substitution rates (average {\tt phyloP} score from UCSC genome browser greater than zero); this approach leaves 61,980 regions with a high density of assayed CpGs that are between 50bp and 758bp long (mean: 95.1bp). The density of fast (in terms of their annotated rate category) CpG sites in these regions ranges from about 10\% to 100\%, with 2,662 sites exceeding 50\%. Figure \ref{fig:browser} shows the longest of the 207 regions with all constituent CpGs in the fast rate category. This 129 bp region with 16 assayed CpG sites overlaps a CpG island, has strong evolutionary sequence conservation across placental mammals, and displays histone modifications correlated with transcriptional enhancer activity in various blood-related cell types~\citep{mouseENCODE}. These attributes suggest a gene regulatory role for this locus.

%===================
\section{Discussion}
%==================

In this manuscript, we adapt phylogenetic Markov models, which are prevalent in comparative genomics and statistical genetics, to accurately and efficiently analyze DNA methylation dynamics across lineage specification. As a proof of concept, we model RRBS methylation data collected from 13 related stages of blood cell development~\citep{bock12}. Using our model, we illustrate that \emph{(i)} CpG site methylation status can be accurately reconstructed using data from related cell types at the same site; \emph{(ii)} the single CpG site resolution of our methylation dynamics estimates enable the discovery of attributes, such as DNA sequence context, that correlate with CpG methylation dynamics; and \emph{(iii)} our models facilitate the identification of genomic regions with highly variable CpG methylation states that are likely functional. 

There are many additional methodologies that could be mapped from the rich reservoir of statistical genomics to the application of modeling methylation dynamics. It will be exciting to see which will prove most useful as genome-wide methylation data continue to be collected to elucidate tissue differentiation and development. 

In Section~\ref{sec:reconstruction}, we demonstrated that progenitor--descendant relationships in the lineage tree can be used to accurately reconstruct the methylation status of CpG sites. This approach is complementary to information provided by other methods for reconstructing DNA methylation that use neighboring CpG sites along the genome~\citep{zhang13}. 
Context-dependent epi-phylogenetic models that integrate such ``vertical'' (i.e., progenitor to descendant) and ``horizontal'' (i.e., distance on chromosome) relationships 

In addition, our strategy is subject to the discretization of methylation status in a population of cells into discrete methylation states. The three states and the thresholds we employ are supported by the distribution of methylation values in our data, but including direct modeling of counts of methylated vs.~unmethylated instances of a CpG site~\citep{ziller13} into our approach could be a promising further direction.

Finally, DNA methylation is just one of several dynamic epigenetic/biochemical modifications regulating precise spatio-temporal gene expression patterns that are essential for proper development. The ``phylo-epigenetic'' approach we have de\-mon\-stra\-ted here provides an integrative, multivariate framework for modeling any epigenetic changes across multiple cell types and lineages.  As phylogenetic models proved essential in the identification and interpretation of functional DNA sequence regions, we believe that phylo-epigenetic models can play a similar role in developing a deeper understanding of epigenetic phenomena and their roles in tissue differentiation and vertebrate development.

\section*{Acknowledgement}

\emph{\bf Funding:} JAC was supported by institutional funds from Vanderbilt University, DK was supported by institutional funds from the University of Pittsburgh School of Medicine.

\bibliographystyle{natbib}
\bibliography{phy-epi}

\end{document}